\documentclass[aps,prb,twocolumn,amsmath,amssymb,superscriptaddress,floatfix]{revtex4}
\usepackage{graphicx}
\usepackage{amssymb}
\usepackage{natbib}
\usepackage{color}
\usepackage{float}
\usepackage{placeins}

\newcommand{\beq}{\begin{eqnarray}}
\newcommand{\eeq}{\end{eqnarray}}

\newcommand{\BFCA}{Ba(Fe$_{1-x}$Co$_{x})_2$As$_{2}$}
\newcommand{\BFCAA}{Ba(Fe$_{0.952}$Co$_{0.048}$)$_{2}$As$_{2}$}
\newcommand{\BFCAAA}{Ba(Fe$_{0.946}$Co$_{0.054}$)$_{2}$As$_{2}$}

\begin{document}

\title{Spin fluctuation anisotropy as a probe of orbital-selective
hole-electron quasiparticle excitations in detwinned {\BFCA}}

\author{Long Tian}
\author{Panpan Liu}
\author{Zhuang Xu}
\affiliation{Center for Advanced Quantum Studies and Department of Physics, Beijing Normal University, Beijing 100875, China}

\author{Yu Li}
\affiliation{Department of Physics and Astronomy, Rice University, Houston, Texas 77005, USA}

\author{Zhilun Lu}
\affiliation{Helmholtz-Zentrum Berlin f{\"u}r Materialien und Energie GmbH, Berlin 14109, Germany}

\author{H. C. Walker}
\affiliation{ISIS Facility, Rutherford Appleton Laboratory, Chilton, Didcot, Oxfordshire OX11 0QX, UK}

\author{U. Stuhr}
\affiliation{Laboratory for Neutron Scattering and Imaging, PSI, CH-5232 Villigen, Switzerland}

\author{Guotai Tan}
\author{Xingye Lu}
\affiliation{Center for Advanced Quantum Studies and Department of Physics, Beijing Normal University, Beijing 100875, China}

\author{Pengcheng Dai}
\email{pdai@rice.edu}
\affiliation{Department of Physics and Astronomy, Rice University, Houston, Texas 77005, USA}
\affiliation{Center for Advanced Quantum Studies and Department of Physics, Beijing Normal University, Beijing 100875, China}

\date{\today}

\begin{abstract}
We use inelastic neutron scattering to study spin excitation anisotropy in mechanically detwinned  \BFCA\  with $x=0.048$  and 0.054. Both samples exhibit a
tetragonal-to-orthorhombic structural transition at $T_s$, a collinear static
antiferromagnetic (AF) order at wave vector ${\bf Q}_1=
{\bf Q}_{\rm AF}=(1,0)$ below the N$\rm \acute{e}$el temperature
$T_N$, and superconductivity below $T_c$ ($T_s>T_N>T_c$).
In the high temperature paramagnetic tetragonal phase ($T\gg T_s$), spin excitations
centered at ${\bf Q}_1$ and ${\bf Q}_2=(0,1)$ are gapless and have four-fold ($C_4$)
rotational symmetry. On cooling to below $T_N$ but above $T_c$, spin excitations become highly anisotropic,
developing a gap at ${\bf Q}_2$ but still are gapless at ${\bf Q}_1$.
Upon entering into the superconducting state, a neutron spin resonance appears at ${\bf Q}_1$ with no magnetic scattering at ${\bf Q}_2$. By comparing these results with those from angle resolved photoemission spectroscopy experiments, we conclude that
the anisotropic shift of the $d_{yz}$ and $d_{xz}$ bands in detwinned \BFCA\ below $T_s$ is associated with the spin excitation anisotropy, and the
superconductivity-induced resonance arises from the electron-hole Fermi surface nesting
of quasiparticles with the $d_{yz}$ orbital characters.
\end{abstract}

\maketitle

\section{introduction}

Unconventional superconductors such as copper oxides, iron pnictides, and heavy Fermions are interesting because superconductivity
in these materials is derived from their long-range antiferromagnetic (AF) ordered parent compounds \cite{scalapinormp}.  Although there is no consensus on the microscopic origin of superconductivity, there is increasing evidence that electron pairing in these superconductors is
mediated by spin fluctuations (excitations)
\cite{scalapinormp,Fradkin,Eschrig,dairmp,Kenzelmann}.  In particular,
superconductivity is intertwined with magnetic degrees of freedom, and forms a
 state coexisting with the static AF order in the underdoped regime \cite{scalapinormp,Fradkin,Eschrig,dairmp,Kenzelmann}.
Therefore, to understand
the fundamental interactions that lead to unconventional
superconductivity, it is important to investigate how magnetism interacts with superconductivity
in the coexisting regime of unconventional superconductors.

\begin{figure}[t]
\includegraphics[scale=0.25]{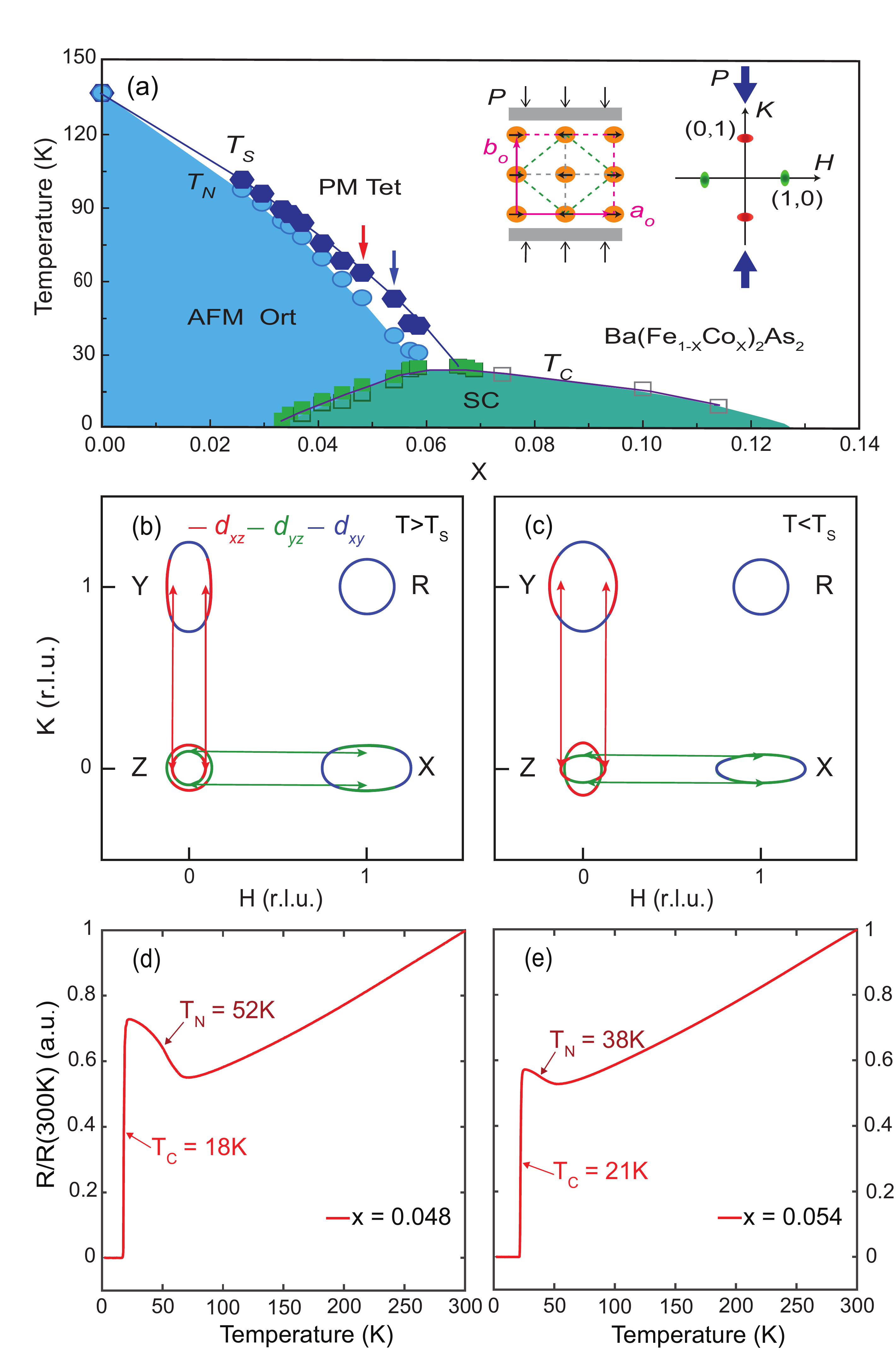}
\caption{ (a) The phase diagram of {\BFCA} with the arrows indicating the Co-doping concentrations of our samples ($x = 0.048$ and $0.054$). The left hand inset shows a schematic of
the collinear AF ordering of the Fe spins in real space and the applied uniaxial pressure direction
is marked by vertical arrows.
The right hand inset is the corresponding reciprocal space map showing the AF ordering
wave vector ${\bf Q}_1$ (green) and wave vector ${\bf Q}_2$ (red).
The filled hexagonal and circular points represent $T_s$ and $T_N$, respectively, obtained from resistance measurements. The filled and open square points mark $T_c$ determined
from the resistance and magnetic susceptibility measurements, and from Refs. \cite{Nandi}.  Schematic Fermi surfaces of underdoped {\BFCA} in (b) paramagnetic tetragonal state and (c) nematic
state below $T_s$. The red, green, and blue colors represent $d_{xz}$, $d_{yz}$, and $d_{xy}$ orbitals,
respectively. The arrows mark nesting wave vectors ${\bf Q}_1$ and ${\bf Q}_2$
between the $Z$ and $X/Y$ points. Our definition of the $X/Y$ is switched from that of
the ARPES work \cite{Pfau2019}.
The normalized temperature dependent resistance data indicates the superconducting transition temperatures of (d) $x = 0.048$ and (e) $x = 0.054$.}
\end{figure}

In the case of Co-underdoped iron pinictide superconductors such as {\BFCA}
with $0.03< x < 0.065$, they exhibit a
tetragonal-to-orthorhombic structural transition at $T_s$, a collinear static
AF order below the N$\rm \acute{e}$el temperature $T_N$, and superconductivity below $T_c$ ($T_s>T_N>T_c$)
as shown in Fig. 1(a) \cite{stewart,Ni08PRB,Lester,Pratt,Christianson,Nandi}. As a function of decreasing temperature, a collinear static AF order is established below $T_N$ at wave vector ${\bf Q}_1=
{\bf Q}_{\rm AF}=(1,0)$ [inset in Fig. 1(a)]. On further cooling across $T_c$, the static ordered moment decreases below $T_c$ accompanied by the formation of a
neutron spin resonance coupled to superconductivity \cite{Pratt,Christianson,Nandi,Luo13}.
  For optimally and overdoped
{\BFCA}, where their $T_s$ and $T_N$ are suppressed and
the system is in the paramagnetc tetragonal state, neutron spin resonance
occurs at wave vectors ${\bf Q}_1=(1,0)$ and ${\bf Q}_2=(0,1)$,
and therefore obeys fourfold rotational
 ($C_4$) symmetry of the underlying tetragonal lattice \cite{Lumsden09,Schi09,SLi09,MWang2013}.
While superconductivity clearly competes with static AF order
in underdoped {\BFCA} \cite{Pratt,Christianson,Nandi}, much is unclear concerning how the superconductivity-induced neutron spin resonance
interacts with spin waves from the AF ordered phase. Although the collinear AF order and associated
low energy spin waves
should have two-fold rotational ($C_2$) symmetry below $T_N$, the observed
resonance and spin waves have the $C_4$ symmetry from the presence of twin domains of the orthorhombic phase below $T_s$ \cite{Pratt,Christianson,Nandi}. Therefore, to understand the interplay between spin waves associated with static AF order and the neutron spin resonance connected with superconductivity, one must carry out inelastic neutron scattering experiments on detwinned samples with static AF order and superconductivity.

\begin{figure}[t]
\includegraphics[scale=.42]{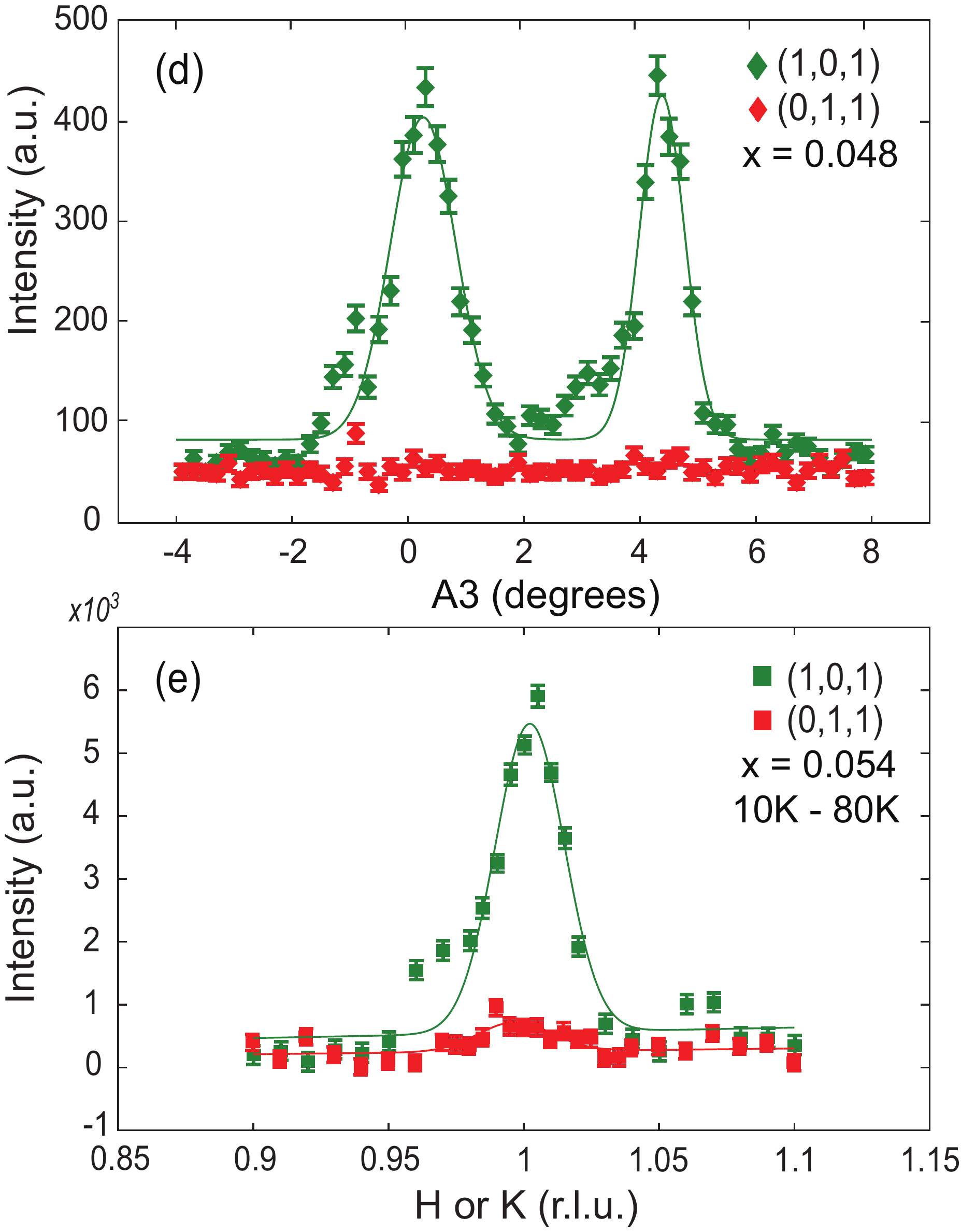}
\caption{ (a) Elastic Rocking curve scans of the sample angle ($A3$) around ${\bf Q}_1=(1,0,1)$
and ${\bf Q}_2=(0,1,1)$ in uniaxial strained Ba(Fe$_{0.952}$Co$_{0.048}$)$_{2}$As$_{2}$. The
double peaks show the sample has two major domains separated by about $4.5$ degrees. (b) Temperature differences of the transverse scans around ${\bf Q}_1$
and ${\bf Q}_2$ in unaixial strained Ba(Fe$_{0.946}$Co$_{0.054}$)$_{2}$As$_{2}$. }
\end{figure}

In this article, we report comprehensive inelastic neutrons scattering experiments designed to study spin excitations in detwinned {\BFCA} with coexisting AF order and superconductivity. From the phase diagram of {\BFCA} in Figure 1(a) established from our own and
previous transport work \cite{Nandi}, we know
that static AF order
decreases with increasing Co-doping, and
competes with superconductivity,
which increases with increasing Co-doping \cite{Pratt,Christianson,Nandi}.
To study the interplay of spin waves associated with the static AF order and resonance connected with superconductivity, one must judiciously choose the Co-doping concentrations where the strength of the spin waves are comparable with the superconductivity-induced resonance \cite{MWang2016,HQLuo2013}. For this purpose, we prepared single crystals of {\BFCA} with $x = 0.048$ and 0.054 \cite{Ni08PRB,Lester,Pratt}.  At zero external uniaxial pressure \cite{Dhital2012,Dhital2014,YSong2013,Tam2017}, the $x = 0.048$ samples have $T_c=18$ K, $T_N=52$ K, and $T_s=63$ K [Fig. 1(d)], and 0.054 crystals have
$T_c=21$ K, $T_N=38$ K, and $T_s=53$ K [Fig. 1(e)]. Upon detwinning these crystals using a device similar to previous work with about 40 MPa uniaxial pressure \cite{Lu14,xylu18},
the system no longer has a clean $T_s$ because the $C_4$ rotational symmetry
in the tetragonal phase is already broken by the applied pressure and $T_N$ increases several K under pressure consistent with the earlier
work \cite{Dhital2012,Dhital2014,YSong2013,Tam2017}.
We carried out inelastic neutron scattering measurements on nearly 100\% detwinned {\BFCA}. In the normal state
above $T_c$ but below $T_N$, spin excitations are gapless and increase with increasing energy at the AF ordering wave vector ${\bf Q}_1=(1,0)$,
but have a $\sim$12 meV gap at ${\bf Q}_2=(0,1)$ [see inset in Fig. 1(a)], showing
strong magnetic anisotropy. On cooling to below $T_c$, a neutron spin resonance at $E_r$
and a spin gap at energies below $E_r$ are
formed at ${\bf Q}_1=(1,0)$, but there is no superconductivity-induced
magnetic scattering at ${\bf Q}_2=(0,1)$. On warming to temperatures slightly above the finite pressure $T_N$, the spin excitations are still anisotropic, showing much strong scattering at ${\bf Q}_1=(1,0)$. Finally, on warming to temperatures well above $T_N$ and zero pressure $T_s$, spin excitations become
the same at ${\bf Q}_1=(1,0)$ and ${\bf Q}_2=(0,1)$, and obey the
$C_4$ rotational symmetry.
By comparing these results with those from angle resolved photoemission spectroscopy (ARPES) experiments \cite{MYi2011,YZhang2011,Brouet2012,MYi2017,Pfau2019,Watson2019}, we conclude that
the anisotropic shift of the $d_{yz}$ and $d_{xz}$ bands of the electron Fermi pockets at the
$X$ and $Y$ points in detwinned \BFCA\ below $T_s$ is
associated with the spin excitation anisotropy, and the
superconductivity-induced resonance arises from the electron-hole Fermi surface nesting
of quasiparticles with the $d_{yz}$ orbital characters along the ${\bf Q}_1$ direction
as shown schematically in Fig. 1(c).

\begin{figure}[t]
\includegraphics[scale=.35]{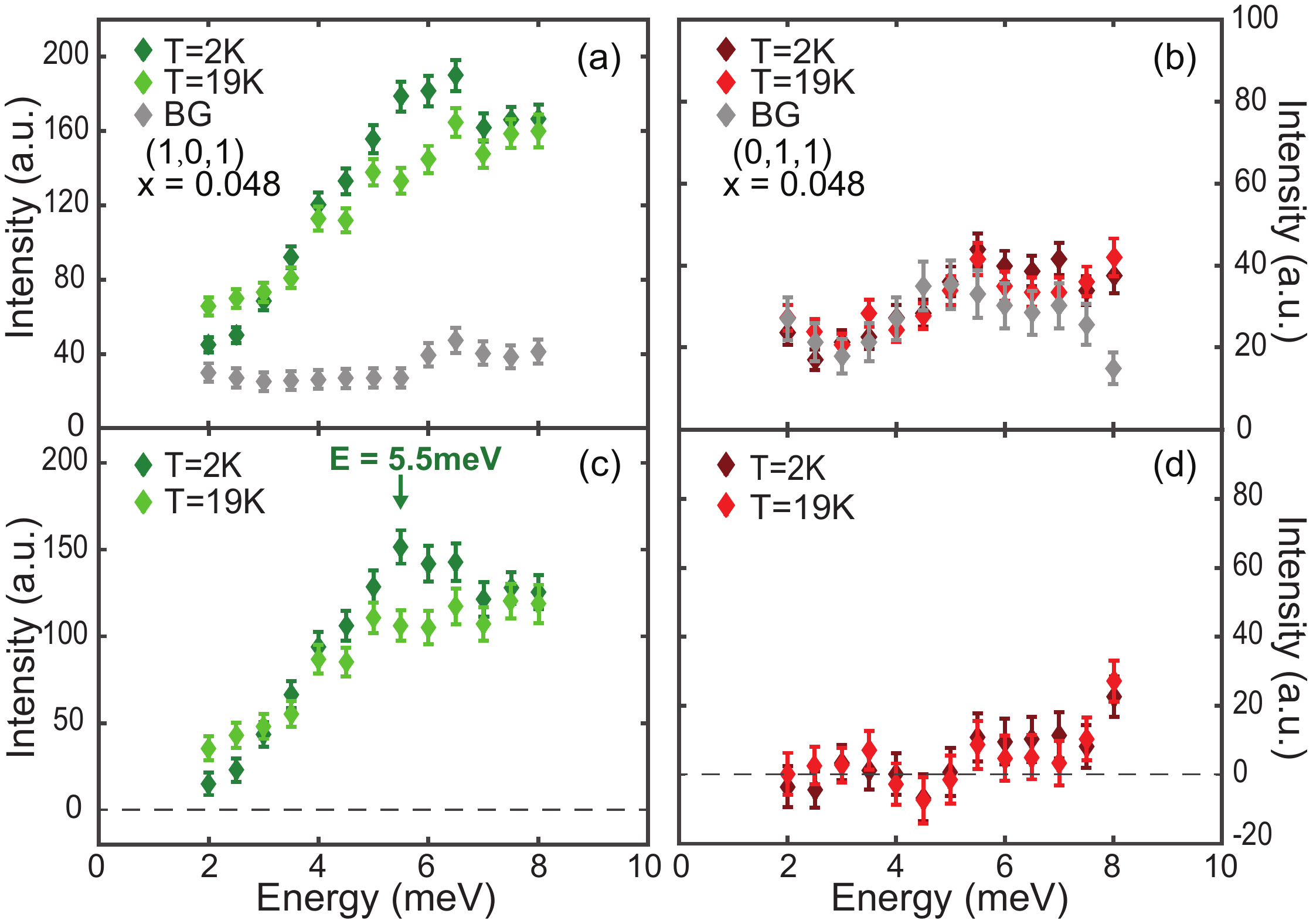}
\caption{ (a) Constant-${\bf Q}$ scans at ${\bf Q}_1$ below and above $T_c$ in {\BFCAA}. The resonance is seen as intensity gain below $T_c$ around $E_r\approx 5.5$ meV.
The gray data points represent the background scattering. (b) Identical scans at ${\bf Q}_2$.
(c,d) Background subtracted constant-${\bf Q}$ scans below and above $T_c$ at
the ${\bf Q}_1$ and ${\bf Q}_2$ points. The vertical arrow indicates
the position of the neutron spin resonance.}
\end{figure}

\section{Experimental Results}

Our neutron scattering experiments were carried out on the FLEXX cold neutron three-axis spectrometer at Helmholtz Zentrum Berlin, Germany, and the MERLIN neutron time-of-flight (TOF) chopper spectrometer at ISIS, Rutherford-Appleton Laboratory, UK \cite{ISIS}.
The detwinning ratio for {\BFCAAA} samples used at ISIS was measured on the EIGER thermal neutron three-axis spectrometer at Paul Scherrer Institute, Switzerland.
Sizable single crystals of {\BFCA} were grown by self-flux method and cut along the $a$ and $b$ axes directions of the orthorhombic lattice below $T_s$ \cite{dairmp}.  Each cut sample was mounted on a specially designed aluminum-based sample holder with uniaxial pressure applied along the $b$-axis
direction \cite{Lu14,xylu18}. The total mass of our samples is $\sim$2.5 g for the $x = 0.048$ used for FLEXX experiment, $\sim$3.6 g for the $x = 0.054$ used for MERLIN and EIGER experiments. The momentum transfer ${\bf Q}$ in three-dimensional reciprocal
space is defined as
${\bf Q}$ = $H{\bf a}^* + K{\bf b}^* + L{\bf c}^*$, where $H, K$ and $L$ are Miller indices and ${\bf a}^* = \hat {\bf a}2\pi/a$, ${\bf b}^* = \hat {\bf b}2\pi/b$, ${\bf c}^* = \hat {\bf c}2\pi/c$ with $a$ = 5.615 \AA, $b$ = 5.573 \AA\ and $c$ = 12.95 \AA\ in the low-temperature orthorhombic state \cite{Pratt,Christianson,Nandi}.
In this notation, the AF order occurs at the in-plane wave vector ${\bf Q}_{\rm AF}=(1,0)$,
and there should be no elastic magnetic scattering at wave vector $(0,1)$.
For measurements on three-axis spectrometers, we aligned the samples in the $[1,0,1] \times [0,1,1]$ scattering plane where we can measure the static magnetic order and excitations at both ${\bf Q}_{\rm AF}={\bf Q}_1 = (1,0,1)$ and ${\bf Q}_2 = (0,1,1)$ simultaneously \cite{Lu14}.
The fixed final neutron energies are $E_f=5$ and 14.7 meV for FLEXX and EIGER experiments,  respectively. For experiments on TOF spectrometer MERLIN, the direction of the incident beam is parallel to the $c$ axis. Using multi-$E_i$ mode with a primary incident neutron beam energy of
$E_i = 80$ meV and Fermi chopper frequency of $\omega = 250$ Hz,
we were able to measure with two additional incident energies
of $25$ and $12$ meV, thus allowing spin excitations up to
$E<70$ meV to be probed.

\begin{figure}[t]
\includegraphics[scale=.5]{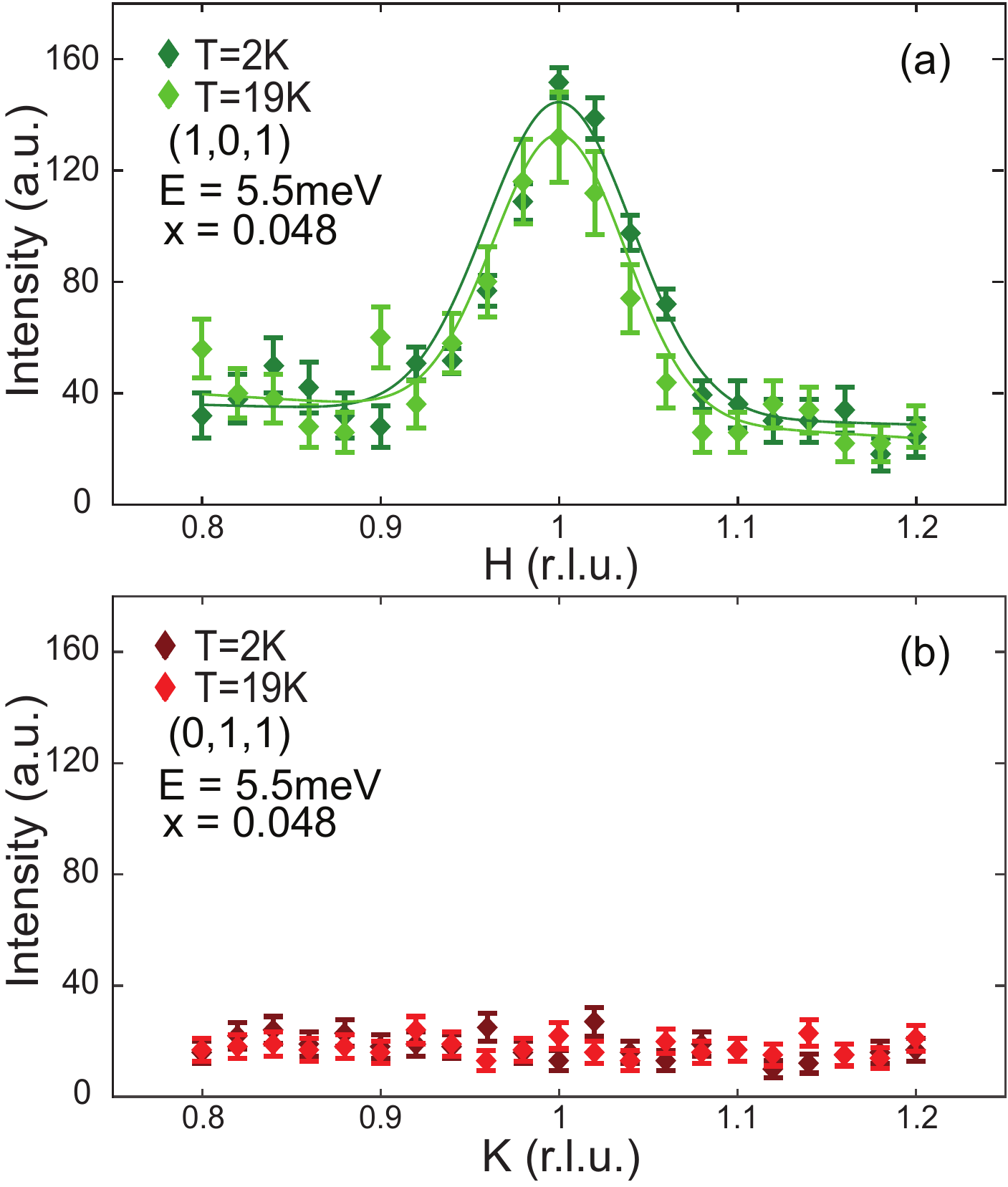}
\caption{ Constant-energy scans in {\BFCAA} below and above ${T_c}$ with $E_r = 5.5$
meV along the $[H,0,1]$ and $[0,K,1]$ directions
around (a) ${\bf Q}_1$ and (b) ${\bf Q}_2$.}
\end{figure}

Figure 1(a) shows the phase diagram of electron doped {\BFCA}
as determined from our transport measurements. Consistent with previous work \cite{Ni08PRB},
we find that the ratio between the actual and nominal Co-doping level is about $0.74$.
For the experiments, we chose {\BFCA} with Co-doping levels $x = 0.048$ and $0.054$
as marked by vertical arrows in Fig. 1(a). Figures 1(b) and 1(c) show Fermi surfaces
of underdoped {\BFCA} with coexisting AF order and superconductivity
above and below the zero pressure $T_s$, respectively,
as obtained from ARPES experiments on uniaxial pressure detwinned samples \cite{MYi2011,YZhang2011,Brouet2012,MYi2017,Pfau2019,Watson2019}.
The temperature dependence of the normalized resistance for
{\BFCAA} and {\BFCAAA} reveals superconducting
transition temperatures of $T_c=18$ K and 21 K, respectively [Figs. 1(d) and 1(e)].

\begin{figure}[htbp!]
\includegraphics[width=7.0cm]{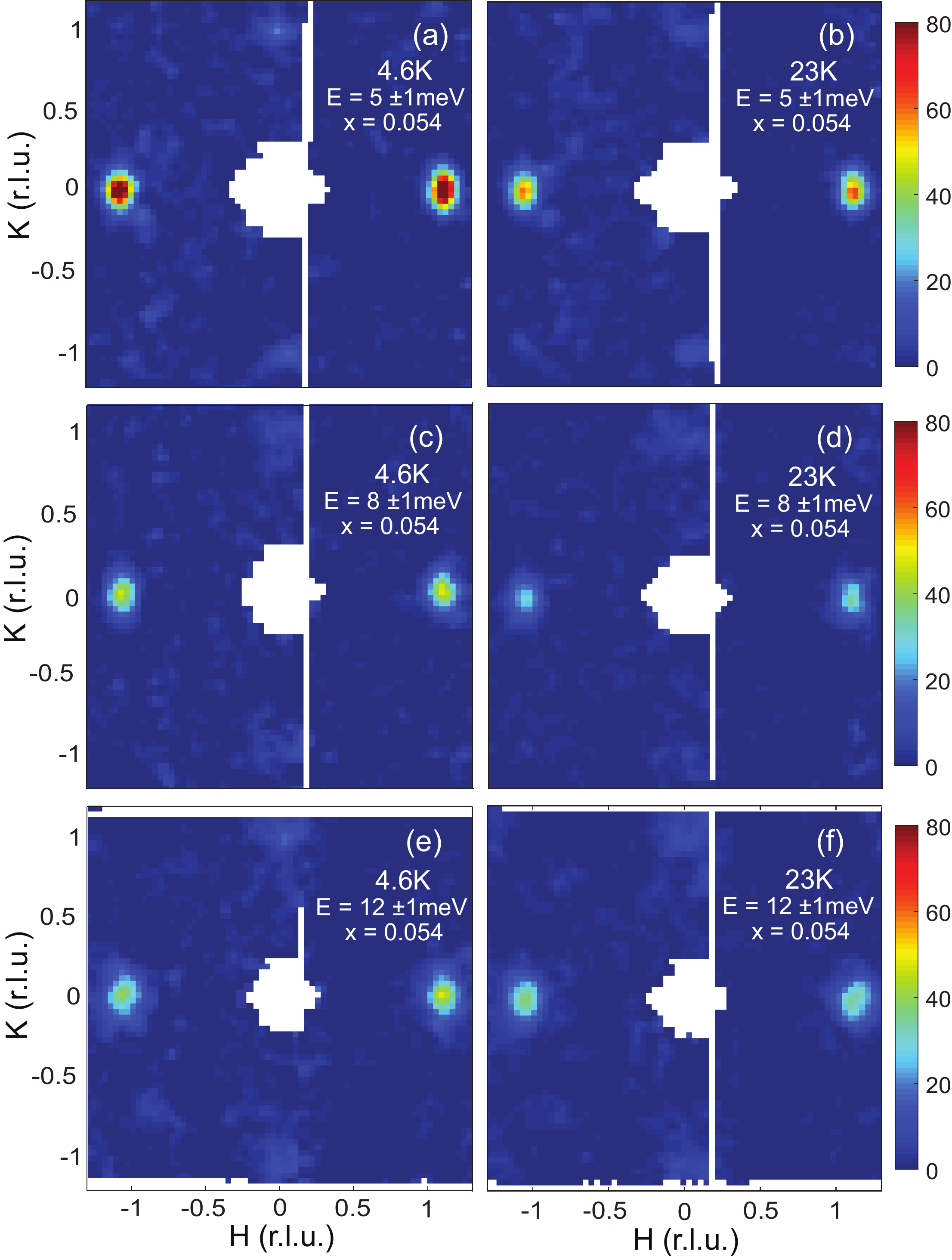}
\caption{Two-dimensional images of the spin excitations of {\BFCAAA} within the $[H,K]$ plane
below and above $T_c$ at energies of (a,b) $E = 5 \pm 1$ meV; (c,d) $E = 8 \pm 1$ meV; (e,f)
$E = 12 \pm 1$ meV. The vertical color bars indicate scattering intensity in arbitrary units.}
\end{figure}

In order to carry out inelastic neutron scattering experiments on detwinned
{\BFCAA} and {\BFCAAA}, one must mount crystals in a uniaxial detwinning device and
apply uniaxial pressure along one axis of the orthorhombic lattice to detwin the samples [see inset of Fig. 1(a)]. For fully detwinned samples, one would expect to observe magnetic Bragg intensity at ${\bf Q}_1$ but no magnetic signal
at ${\bf Q}_2$. Figures 2(a) and 2(b) show the background subtracted rocking curve elastic
scans around ${\bf Q}_1$ and ${\bf Q}_2$ for {\BFCAA} and {\BFCAAA}, respectively.
While the {\BFCAA} sample is $\sim$100\%\ detwinned, there is a weak peak at ${\bf Q}_2 = (0,1,1)$
for {\BFCAAA}. Defining the detwinning ratio as  $\eta = (I_{10} - I_{01})/(I_{10} + I_{01})$, where $I_{10}$ and $I_{01}$ are magnetic scattering at ${\bf Q}_1$ and ${\bf Q}_2$,
respectively \cite{YuSong15PRB,TChen2019}, we find that {\BFCAAA} has a detwinning ratio
of $\eta\approx85$ \%.  Using the measured $\eta$, we can estimate the intrinsic magnetic
scattering at ${\bf Q}_1$ and ${\bf Q}_2$ for different energy transfers, and thus determine the energy dependence of the magnetic scattering at these wave vectors and the related dynamic magnetic
susceptibility $\chi^{\prime\prime}({\bf Q},E)$ \cite{YuSong15PRB,TChen2019}.

\begin{figure*}[t]
\includegraphics[scale=.4]{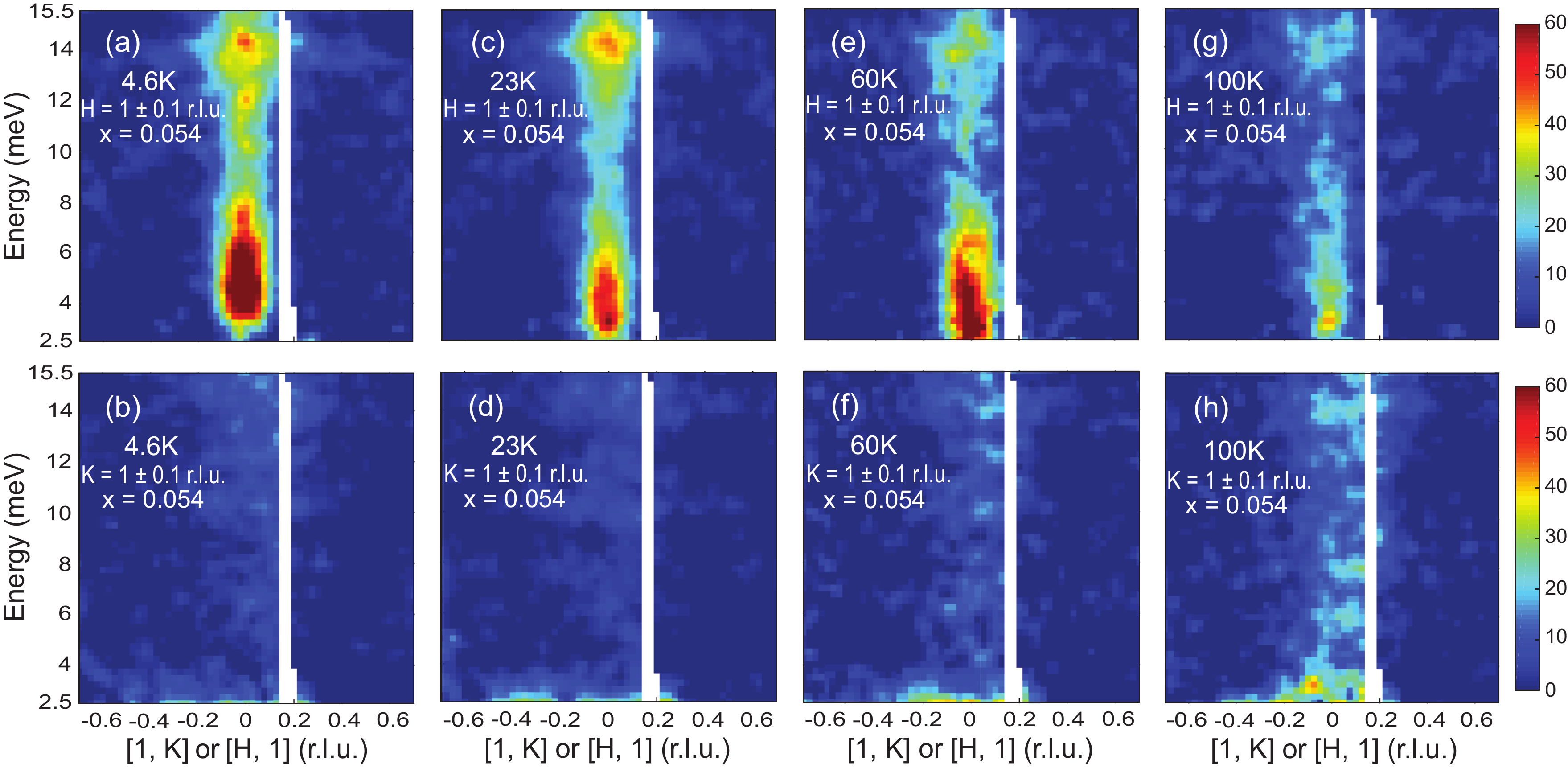}
\caption{Two-dimensional images of magnetic scattering in {\BFCAAA}
along the $[1,K]$ and $[H,1]$ directions  at temperatures (a,b) 4.6 K; (c,d) 23 K; (e,f) 60 K;
(g,h) 100 K. The incident beam energy is $E_i = 25$ meV and the
partial detwinning ratio of the sample has been corrected for.}
\end{figure*}

We first present our inelastic neutron scattering results for {\BFCAA}. Figures 3(a) and 3(b) show the constant-{\bf Q} scans at ${\bf Q}_1$ and ${\bf Q}_2$ below and above $T_c$, as well as background scattering at wave vectors transversely rotated $\sim$15$^\circ$ from ${\bf Q}_1$ and ${\bf Q}_2$. At ${\bf Q}_1$, the scattering  increases with increasing energy, and superconductivity induces a resonance at $E_r\approx 5.5$ meV
below $T_c$ [Fig. 3(a)]. Figure 3(c) shows background subtracted scattering, suggesting that superconductivity
opens a spin gap below about 2 meV.  At ${\bf Q}_2$, we find no discernible signal above
the background scattering both below and above $T_c$, suggesting that the presence of a large spin gap
in the normal state and superconductivity does not induce
any magnetic intensity [Fig. 3(b)]. The background subtracted scattering in Fig. 3(d) suggests the presence of a spin gap of $\sim$7 meV in both the normal and superconducting states, with detectable magnetic scattering above 8 meV.
To further confirm these results, we show in Figs. 4(a) and 4(b) constant-energy scans at $E_r=5.5$ meV at
${\bf Q}_1$ and ${\bf Q}_2$, respectively. While there is a clear magnetic peak centered at
${\bf Q}_1$ that enhances the intensity below $T_c$, no discernible peak is seen at ${\bf Q}_2$ below or above $T_c$.

\begin{figure}[t]
\includegraphics[scale=.4]{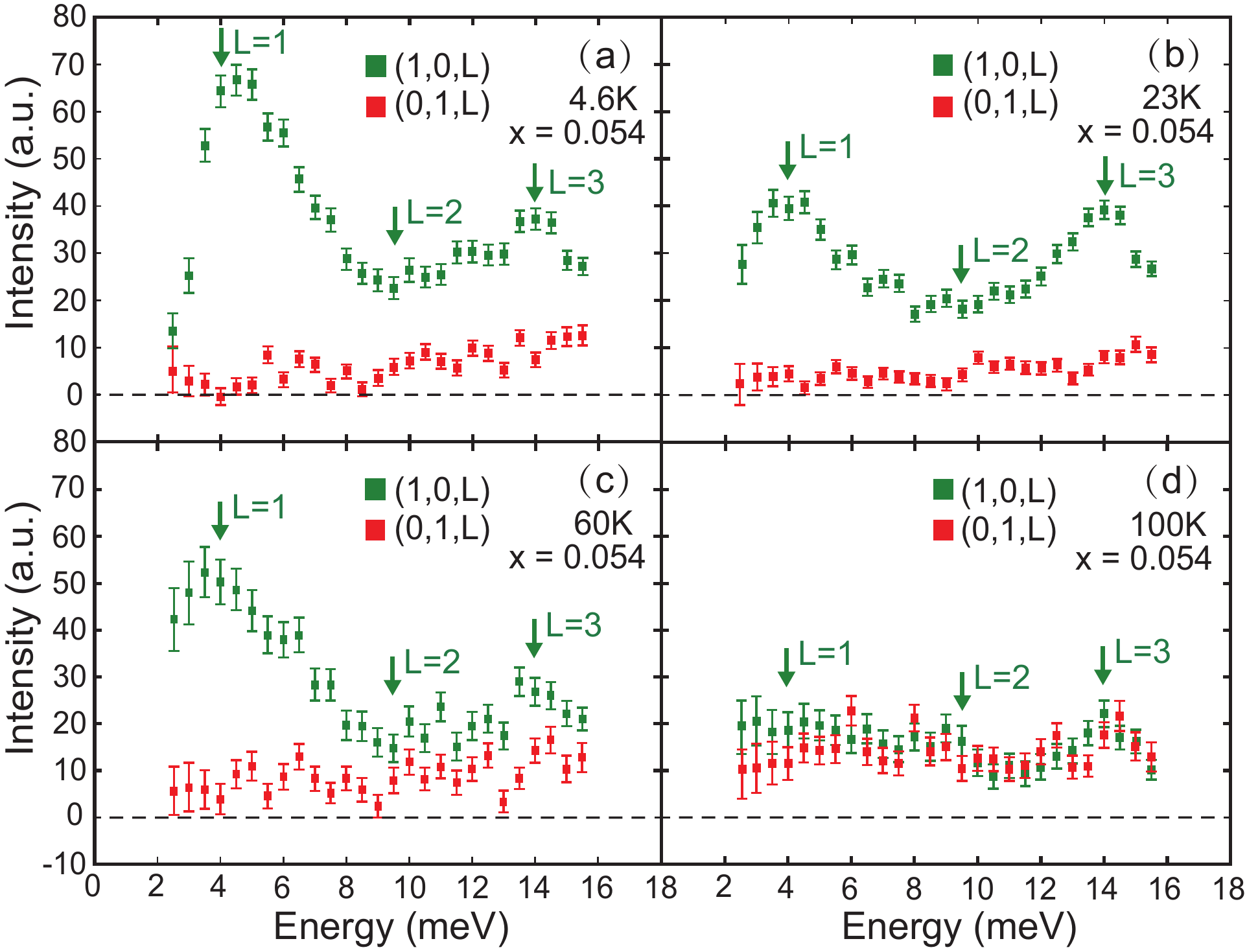}
\caption{The energy/$c$-axis wave vector dependence of the spin excitations around the in-plane wave vector ${\bf Q}_1$ (green) and ${\bf Q}_2$ (red) positions at temperatures (a) 4.6 K, (b) 23 K, (c) 60 K, (d) 100 K. The incident beam energy is $E_i=25$ meV, and
the vertical arrows indicate the energy values with integer $L$. The partial detwinning ratio
has been corrected for.}
\end{figure}

The two-dimensional (2D) magnetic scattering images of {\BFCAAA} in the $(H,K)$ plane for different energy
transfers
below ($T=4.6$ K) and above ($T=23$ K) $T_c$  are shown in Figures 5(a), (c), (e) and 5(b), (d), (f), respectively.
At $E = 5\pm1$ meV, the scattering is centered around ${\bf Q}_1$ and clearly enhances below $T_c$, and there is no scattering at ${\bf Q}_2$ [Figs. 5(a) and 5(b)]. On increasing energy to $E = 8\pm1$ meV,
the situation is similar at ${\bf Q}_1$ but there may be some scattering at ${\bf Q}_2$ [Figs. 5(c) and 5(d)]. On increasing energy to $E = 12\pm1$ meV, superconductivity
has little effect on spin excitations at ${\bf Q}_1$ and there is
weak magnetic signal at ${\bf Q}_2$ [Figs. 5(e) and 5(f)].

\begin{figure*}[t]
\includegraphics[scale=.3]{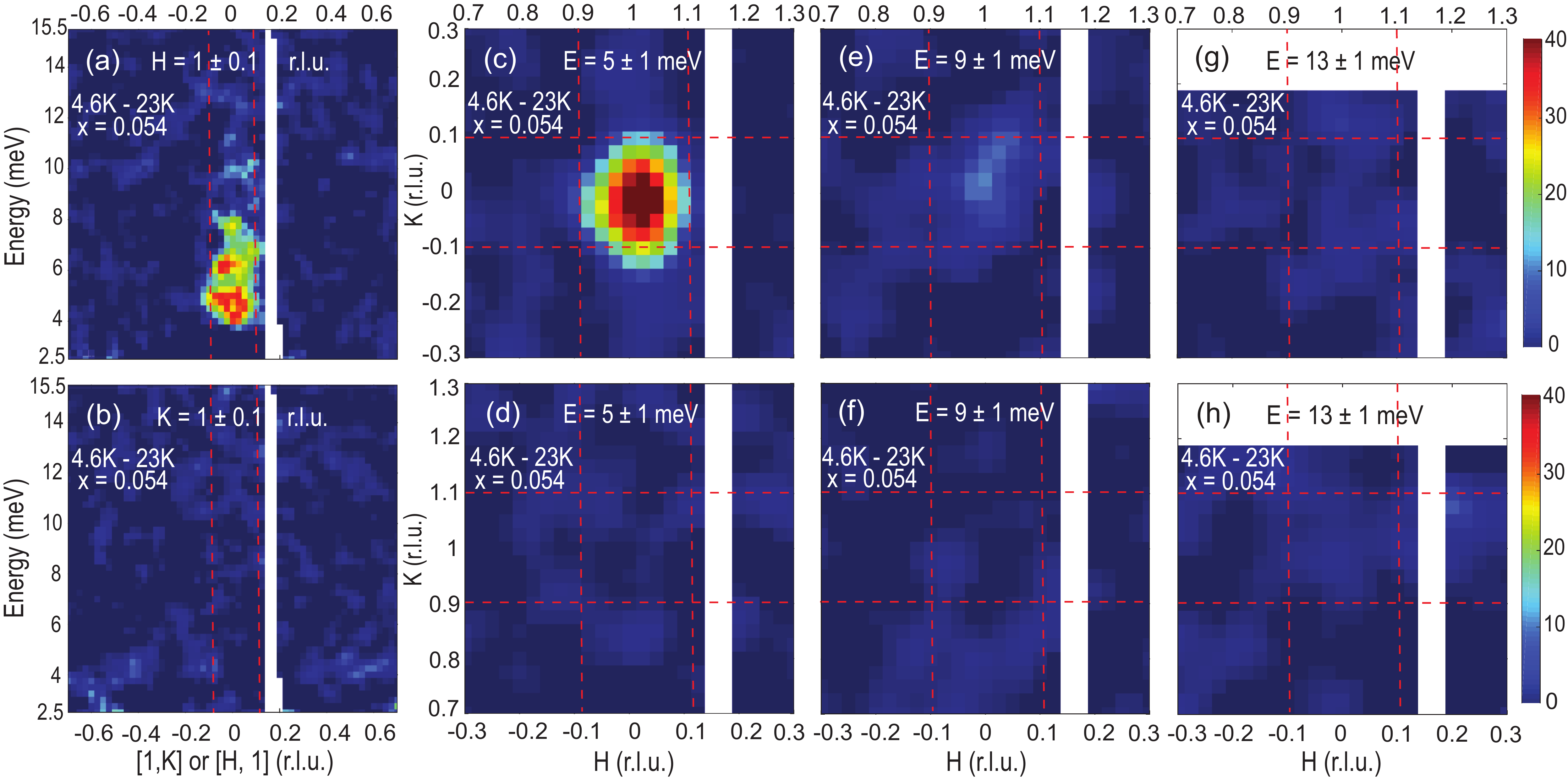}
\caption{(a,b) The energy dependence of the resonance along the ${\bf Q} = [1,K]$ and $[H,1]$ directions in {\BFCAAA}. (c-h) The wave vector dependence of the spin resonance at $E = 5 \pm 1$ meV, $9 \pm 1$ meV, $13 \pm 1$ meV around ${\bf Q}_1$. The vertical and horizontal dashed lines indicate the in-plane momentum integration range used in (a,b).}
\end{figure*}

Figure 6 summarizes the energy dependence of the spin excitations at ${\bf Q}_1$ and ${\bf Q}_2$ as a function of increasing temperature. The energy dependence of the scattering is obtained by integrating wave vectors
 $0.9 < H < 1.1$ around ${\bf Q}_1$ in Figs. 6(a), (c), (e), (g)  and $0.9 < K < 1.1$ in Figs. 6(b), (d), (f), (h) around ${\bf Q}_2$. The effect of the partial detwinning ratio was corrected using the method developed in Ref. \cite{TChen2019}. Consistent with Figs. 4 and 5, we find that superconductivity
induces a broad resonance
(or two resonances) around $E_r\approx 5$ meV at ${\bf Q}_1$ [Figs. 6(a) and 6(c)] \cite{MWang2016,CZhang2013,PSteffens2013,Fwaber2017,Chenglin16PRB}, but has no effect at ${\bf Q}_2$ [Figs. 6(b) and 6(d)] \cite{WYwang2017}. On warming to 60 K, which is above the zero pressure $T_N$ and $T_s$, there is still
clear magnetic excitation anisotropy [Figs. 6(e) and 6(f)].  Finally, on warming to 100 K, the spin excitations
at ${\bf Q}_1$ and ${\bf Q}_2$ become essentially identical with no observable anisotropy.

To accurately determine the dynamic magnetic susceptibility anisotropy, which is associated with spin nematic order \cite{CFang,CXu,Fernandes2011,Fernandes,Fernandes12,SLiang13,Qimiao2016} and may be important for superconductivity \cite{Metlitski,Lederer}, we cut the data in Fig. 6 along the TOF direction, which couples the energy transfer of the
spin excitations with $L$ modulation. Figures 7(a)-7(d) show the temperature dependence of the magnetic scattering
at  ${\bf Q}_1$ and ${\bf Q}_2$. Consistent with earlier work \cite{Lumsden09,Schi09}, we find that spin excitations
at ${\bf Q}_1$ have a strong $L$ modulation with high magnetic intensity at $L=1, 3, \cdots$ [Figs. 7(a)-7(c)].
The spin excitations have a gap of about 8 meV at ${\bf Q}_2$ below 23 K [Fig. 7(b)], and become gap-less and similar to those at ${\bf Q}_1$ around 100 K [Fig. 7(d)].

\begin{figure}[t]
\includegraphics[scale=.42]{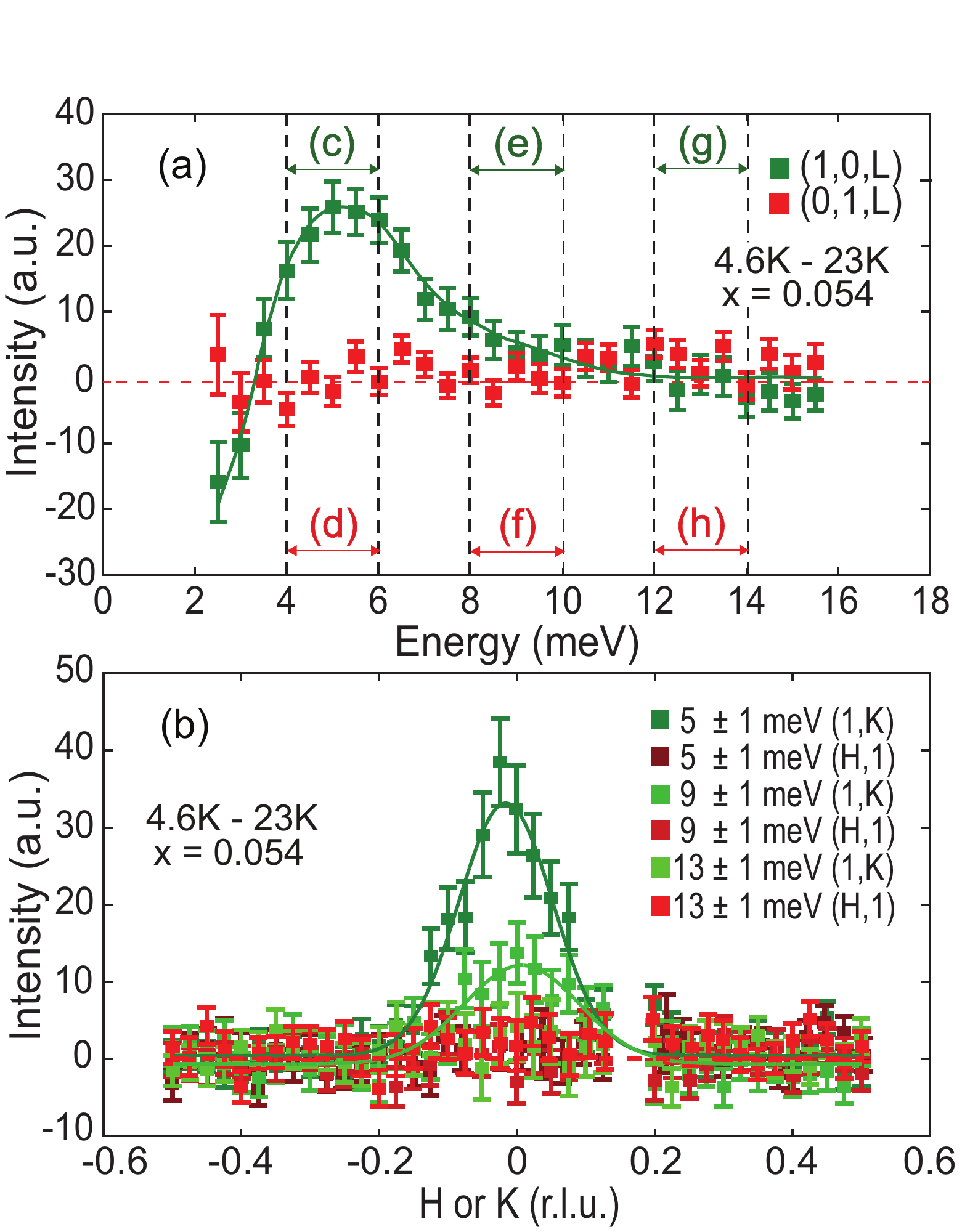}
\caption{(a) Energy dependence of the resonance at ${\bf Q}_1$ and ${\bf Q}_2$ obtained
by taking temperature differences of the constant-${\bf Q}$ cuts below and above $T_c$.
The vertical dashed lines indicate the energy integration range in Figs. 8(c-h).
(b) In-plane wave vector dependence of the magnetic scattering near the resonance energy
around ${\bf Q}_1$ and ${\bf Q}_2$.}
\end{figure}

The impact of superconductivity on the spin excitations at ${\bf Q}_1$ and ${\bf Q}_2$ can be further evaluated by the temperature differences plot below and above $T_c$. Figures 8(a) and 8(b) confirm that superconductivity induces a broad resonance
at ${\bf Q}_1$ and has no effect at ${\bf Q}_2$.  The in-plane wave vector dependence of the
resonance at ${\bf Q}_1$ in Figs. 8(c), (e), (g)
reveals no strong evidence of an incommensurate dispersive resonance
as seen in electron-doped Ba(Fe$_{0.963}$Ni$_{0.037}$)$_2$As$_2$
\cite{MGKim2013} and hole-doped Ba$_{0.67}$K$_{0.33}$(Fe$_{1-x}$Co$_x$)$_2$As$_2$ with
$x=0$ and 0.08 \cite{RZhang2018}, possibly due to the fact that {\BFCAAA} is still not close to optimal superconductivity, or our measurements have insufficient in-plane wave vector resolution to resolve the expected transverse incommensurate scattering \cite{MGKim2013}.
Figure 9(a) shows the energy dependence of the resonance at ${\bf Q}_1$ and ${\bf Q}_2$, confirming the
results of Figs. 8(a) and 8(b).  The wave vector dependence of the resonance at different energies
is plotted in Fig. 9(b), which again reveals no evidence of incommensurate scattering.

In previous work on nearly optimally doped BaFe$_{1.9}$Ni$_{0.1}$As$_2$ \cite{YuSong15PRB}
and BaFe$_2$As$_2$ \cite{xylu18}, spin excitation anisotropy associated with the spin-driven Ising-nematic phase was defined as $\delta=(I_{10}-I_{01})/(I_{10}+I_{01})$, where $I_{10}$ and $I_{01}$ are the magnetic scattering
at ${\bf Q}_1$ and ${\bf Q}_2$, respectively. For
BaFe$_2$As$_2$, the spin excitation anisotropy $\delta$ extends to about $E=180$ meV at 7 K (at zero uniaxial pressure, $T_N\approx 138$ K), reduces to $\sim$120 meV at 145 K (around pressure induced $T_N$) \cite{Tam2017} and $\sim$40 meV at 170 K (well above $T_N$ under pressure), and finally becomes isotropic at 197 K \cite{xylu18}. In the case of BaFe$_{1.9}$Ni$_{0.1}$As$_2$, the spin excitation anisotropy $\delta$ is nonzero below about $\sim$60 meV at 5 K (where $T_N\approx T_s\approx 30\pm 5$ K) and unchanged on warming to 35 K \cite{YuSong15PRB}.
Figure 10 shows the temperature dependence of the magnetic scattering at the resonance energy
of $E_r\approx 5$ meV. On warming from 45 K to 100 K, the scattering profile clearly changes from $C_2$ to $C_4$ symmetric and becomes isotropic.

\begin{figure}[t]
\includegraphics[scale=.35]{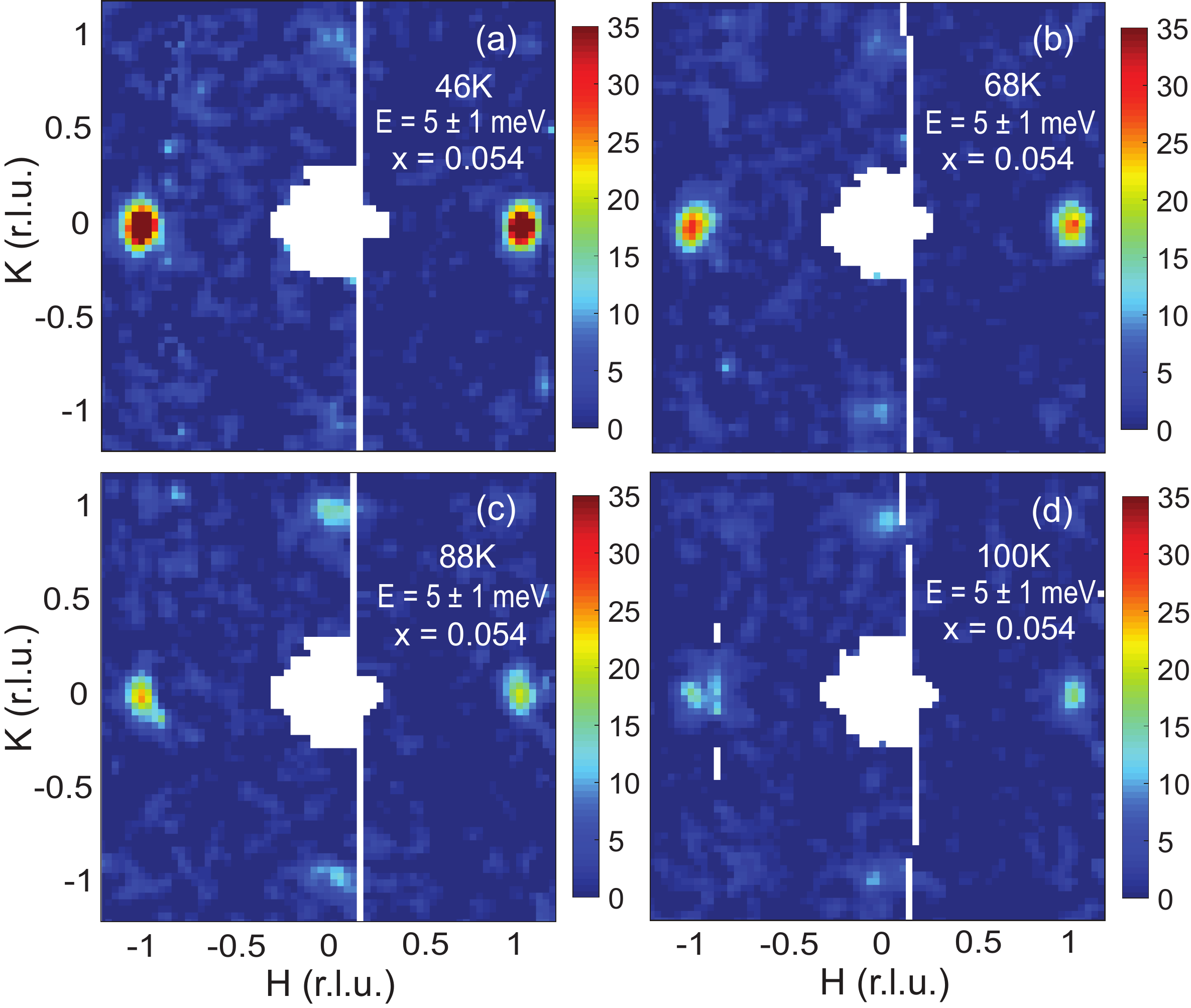}
\caption{Temperature dependence of the $E = 5$ meV spin excitations
in {\BFCAAA} measured with $E_i$ = 25 meV at (a) 46 K, (b) 60 K, (c) 88 K, (d) 100 K.
The scattering intensity has been corrected for the Bose population
factor to obtain $\chi^{\prime\prime}({\bf Q},E)$. }
\end{figure}

To quantitatively determine the
temperature/energy dependence of the spin excitation anisotropy $\delta$
in {\BFCAAA}, we estimate
the energy dependence of the dynamic susceptibility $\chi^{\prime\prime}({\bf Q},E)$,
which can be calculated by $\chi^{\prime\prime}({\bf Q},E)\propto (1-e^{-E/k_BT})I({\bf Q},E)$, at
${\bf Q}_1$ and ${\bf Q}_2$ as a function of increasing temperature. Figure 11(a) shows the energy dependence of
$\chi^{\prime\prime}({\bf Q}_1,E)$ and $\chi^{\prime\prime}({\bf Q}_2,E)$ at $T=4.5$ K, revealing magnetic anisotropy below about 30 meV. Each point of $\chi^{\prime\prime}({\bf Q},E)$ is obtained
by integrating the magnetic scattering over wave vectors $-0.05 < H < 0.05$ and $-0.05 < K < 0.05$ around
${\bf Q}_1$ or ${\bf Q}_2$. On warming to $T=23$ K [Fig. 11(b)], 46 K [Fig. 11(c)], 60 K [Fig. 11(d)],
 78 K [Fig. 11(e)], and 100 K [Fig. 11(f)], the spin excitation anisotropy gradually
decreases and finally vanishes at 100 K.
Figure 11(g) shows the energy dependence of the spin excitation anisotropy
$\delta$ at different temperatures.  Since {\BFCAAA} has $T_N\approx 38$ K similar to nearly optimal electron-doped BaFe$_{1.9}$Ni$_{0.1}$As$_2$ with $T_N\approx 30$ K \cite{YuSong15PRB}, one would expect similar
spin excitation anisotropy in {\BFCAAA} and BaFe$_{1.9}$Ni$_{0.1}$As$_2$, as confirmed by comparing Fig. 11(g) and Fig. 4 of Ref. \cite{YuSong15PRB}.  The magnetic anisotropy is clearly present above the pressure-induced $T_N$.

\begin{figure}[htbp!]
\includegraphics[width=8.0cm]{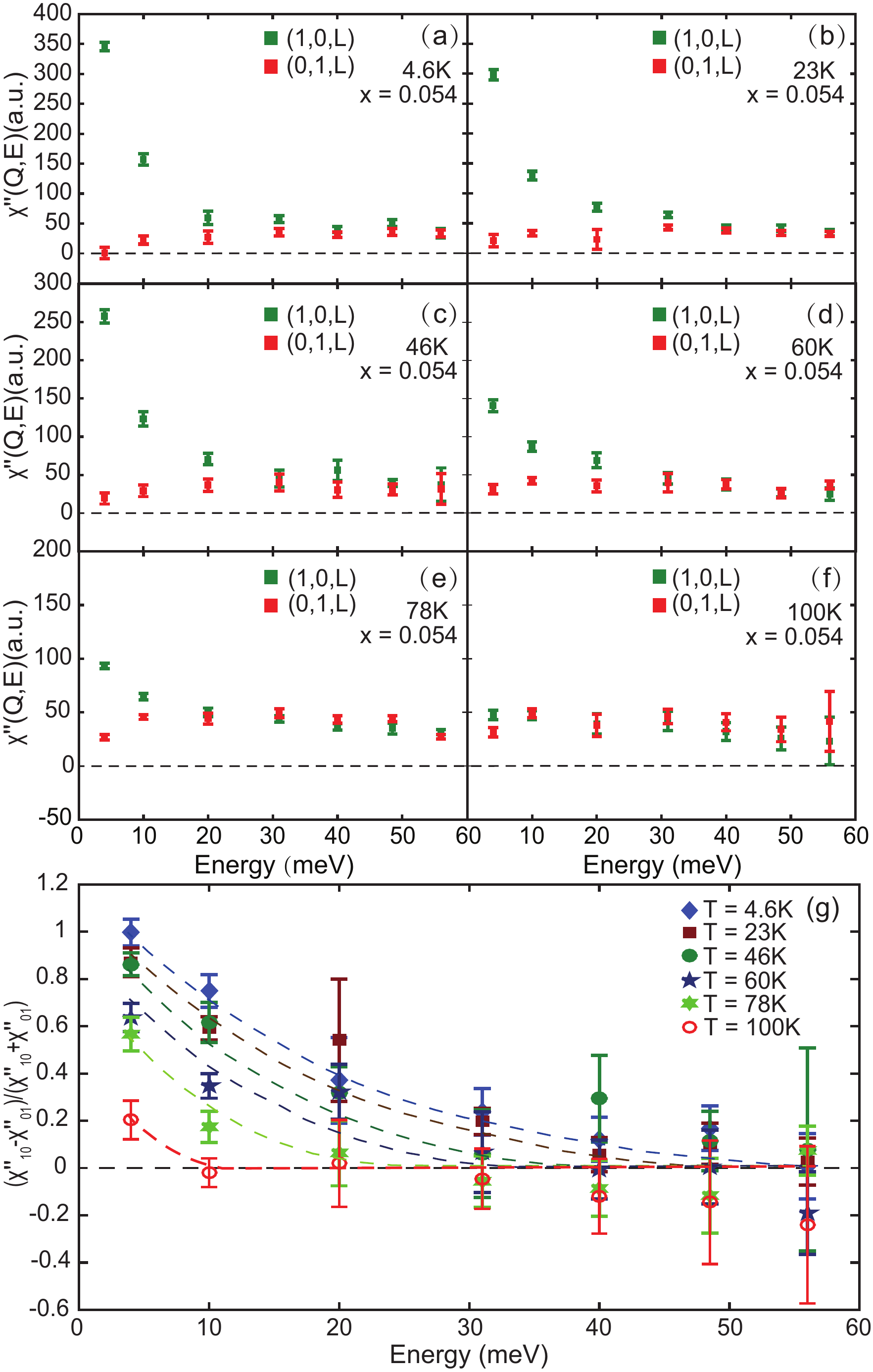}
\caption{Energy dependence of $\chi^{\prime\prime}(E)$,
where the in-plane momentum transfers are integrated around ${\bf Q}_1$ and ${\bf Q}_2$,
 at temperatures (a) 4.6 K, (b) 23 K, (c) 46 K, (d) 60 K, (e) 78 K, (f) 100 K. The first two
data points at low energies in each figure are measured with incident energy $E_i = 25$ meV and the rest are from $E_i = 80$ meV.
The values of $\chi^{\prime\prime}(E)$ are obtained by fitting the transverse cuts
with one Gaussian and linear background.  The Gaussian intensity above background was then
corrected by the magnetic form factor, Bose factor, and the partial detwinning ratio.
(g) Temperature dependence of the spin excitation anisotropy between ${\bf Q_1}$ and ${\bf Q_2}$. The first two data points collected using $E_i = 25$ meV are plotted together.
The dashed lines are guides to the eye.}
\end{figure}

\section{Discussion and Conclusion}

To understand our neutron scattering results in terms of the electron-hole
Fermi surface nesting picture \cite{Hirschfeld2011,Chubukov2012}, we consider electron and hole Fermi surfaces of
uniaixal pressure detwinned {\BFCA} above and below $T_s$ as determined from
ARPES measurements \cite{MYi2011,YZhang2011,Pfau2019,Watson2019}. In the paramagnetic tetragonal state above $T_s$, the hole Fermi surfaces near $\Gamma$ and $Z$ points are composed of $d_{xy}$ and degenerate $d_{xz}/d_{yz}$ orbitals, respectively. If low-energy spin excitations arise from quasiparticle excitations between the hole-electron Fermi surfaces as suggested in an itinerant picture of magnetism and superconductivity
\cite{Hirschfeld2011,Chubukov2012,Dai2012}, the electron-hole Fermi surface nesting of the
$d_{yz}$ and $d_{xz}$ orbital quasiparticles are
along the ${\bf Q}_1$ and ${\bf Q}_2$ directions, respectively. Since a $d_{xy}$ orbital has $C_4$ symmetry,
it cannot by itself induce any anisotropic magnetic scattering
through the hole-electron Fermi surface nesting along the ${\bf Q}_1$ and ${\bf Q}_2$ directions.

When we cool {\BFCA} below $T_s$ in the orthorhombic nematic phase, the $d_{yz}$ band of the electron Fermi surface at $X/Y$ goes up in energy, while the $d_{xz}$ band goes down in energy. Since these are electron pockets, the green part (the $d_{yz}$ band) of the Fermi surface near the Fermi level
will shrink in size, while the red part (the $d_{xz}$ band) of the Fermi surface will expand, resulting in different shaped Fermi surfaces as shown in Fig. 1(c).
At the $Z$ point, which has hole pockets, the changes for the $d_{yz}$ and $d_{xz}$ bands are opposite with much smaller amplitude \cite{Pfau2019,Watson2019}. So we can basically assume that the hole-like Fermi surfaces are not modified much below $T_s$.
If spin fluctuations arise
from intraorbital but interband quasiparticle excitations between hole and electron
Ferim surfaces \cite{JHZhang2010}, spin fluctuations along ${\bf Q}_1$ should arise mostly
from the $d_{yz}$ band scattering between the $Z$ and $X$ points.
Similarly, one would expect spin fluctuations along the ${\bf Q}_2$ direction to arise mostly
from the $d_{xz}$ band scattering between the $Z$ and $Y$ points [Figs. 1(b) and 1(c)].
In the high temperature paramagnetic tetragonal phase, the
$d_{yz}$ and $d_{xz}$ orbital Fermi surfaces are degenerate, resulting in identical shapes for
the electron Fermi pockets at the $X$ and $Y$ points, and an isotropic hole Fermi surface at the $Z$ point
[Fig. 1(b)].  The quasiparticle scattering across the hole-electron Fermi pockets along the ${\bf Q}_1$ and ${\bf Q}_2$ directions and associated spin fluctuations therefore have the same scattering intensity and behave identically.

On cooling to below $T_s$, the lifting of the $d_{yz}$ band
makes the electron Fermi pocket at the $X$ point to be better matched with
the $d_{yz}$ orbital in
the hole pocket, and the
 reduction in the $d_{xz}$ band enhances the oval shape of the electron Fermi pocket at $Y$ point
as shown in Fig. 1(c).  At the $Z$ point, the hole Fermi surfaces also change lineshape due to
the rising $d_{xz}$ band and the reduction of the $d_{yz}$ band, but to a much smaller extent compared
with the shifts in Fermi surfaces at the $X/Y$ points [Fig. 1(c)] \cite{Pfau2019,Watson2019}. Therefore, the major effect of the
tetragonal-to-orthorhombic lattice distortion and associated nematic phase is to change the shapes of the electron Fermi pockets at the $X$ and $Y$ points as shown in Fig. 1(c).
From a pure hole-electron Fermi surface nesting
point of view \cite{Hirschfeld2011},  the nesting condition along the
${\bf Q}_1$ direction improves below $T_s$
because of the better matched hole-electron Fermi surfaces of the $d_{yz}$ band [Fig. 1(c)].
On the other hand, the $\sim$30 meV downward shift of the $d_{xz}$ band below $T_s$ at the $Y$ point
enlarges the electron pocket along the ${\bf Q}_1$ direction \cite{MYi2011} and thus
makes the $d_{xz}$-$d_{xz}$ hole-electron Fermi surface nesting along the ${\bf Q}_2$ direction less favorable.

If we assume that low-energy spin fluctuations in {\BFCA} arise from quasiparticle excitations between hole-electron Fermi pockets at the $Z$ and $X/Y$ points, spin fluctuations along the ${\bf Q}_1$ and ${\bf Q}_2$ directions should be sensitive to the nesting condition
associated with the splitting energy between
the $d_{xz}$ and $d_{yz}$ bands below $T_s$, and become $C_4$ rotational symmetric above
$T_s$. Since the splitting energy between the $d_{xz}$ and $d_{yz}$ bands
is around 60 meV in undoped BaFe$_2$As$_2$,
decreases to about 30 meV in {\BFCA} with $x=0.045$, and vanishes around optimal
superconductivity \cite{MYi2011}, the energy scale of the
spin fluctuation anisotropy
$\delta$ should decrease with increasing $x$ and vanish near optimal superconductivity.
This is qualitatively consistent with the spin anisotropy results on BaFe$_2$As$_2$ \cite{xylu18},
BaFe$_{1.9}$Ni$_{0.1}$As$_2$ \cite{YuSong15PRB}, and {\BFCAAA} [Fig. 11(g)], suggesting that
low-energy spin fluctuations at the wave vector ${\bf Q}_1$ have a strong
$d_{yz}$ orbital character and arise from the
$d_{yz}$-$d_{yz}$ hole-electron Fermi surface quasiparticle excitations.  Since superconductivity-induced neutron spin resonance in underdoped {\BFCA} only appears at ${\bf Q}_1$ (Figs. 3, 8, and 9), it is tempting to argue that superconductivity in these materials arises mostly from electrons with
$d_{yz}$ orbital characters \cite{Hirschfeld2016}. However, such a picture is strictly only true within the itinerant model of magnetism and superconductivity in iron pnictides \cite{Hirschfeld2016}.

The above discussion centers on the assumption that energy splitting of the
$d_{xz}$ and $d_{yz}$ bands originates from orbital/nematic ordering below $T_s$.  In a recent ARPES work on BaFe$_2$As$_2$ \cite{Watson2019}, it was argued that
the splitting of the $d_{xz}$ and $d_{yz}$ bands is induced not by orbital/nematic order at $T_s$, but by static AF order occurring at a
temperature $T_N$ just below $T_s$ \cite{mgkim}.  In this scenario, spin fluctuations occurring at ${\bf Q}_1$ in the AF ordered state originate from spin waves of static ordered moments. The presence of a large spin gap at ${\bf Q}_2$ \cite{xylu18} and effective magnetic exchange coupling anisotropy \cite{leland11} can be well-understood by including a biquadratic coupling term in the local moment Heisenberg Hamiltonian \cite{Wysocki2011,Stanek2011,RYu2012}. In the underdoped regime where superconductivity coexists with AF order, the
broad (or double) resonance mode seen in neutron scattering experiments of
twinned iron pnictides \cite{MWang2016,CZhang2013,PSteffens2013,Fwaber2017,Chenglin16PRB} may arise from interacting spin waves with itinerant electrons \cite{Lv2014}. In this picture, the resonance
associated with the AF order should exclusively appear at ${\bf Q}_1$, while the resonance associated with
itinerant electrons and simple nested Fermi surfaces should appear at both ${\bf Q}_1$
and ${\bf Q}_2$ \cite{Lv2014}. Our results in Figs. 8 and 9 clearly disagree with this picture.

Alternatively, the neutron spin resonance \cite{MWang2016,CZhang2013,PSteffens2013,Fwaber2017,Chenglin16PRB} can arise from orbital-selective paring-induced superconducting gap anisotropy \cite{RYu2014}.
Here, the broadening of the resonance is a consequence of anisotropic superconducting gap in the electron pockets at the $X$ and $Y$ points. Below $T_s$, the unfavorable nesting condition of
the $d_{xz}$ band along the ${\bf Q}_2$ means low-energy spin excitations are gapped at ${\bf Q}_2$.
Therefore, the appearance of the resonance exclusively at ${\bf Q}_1$ suggests that superconducting electrons have mostly the $d_{yz}$ orbital characters below the nematic ordering temperature $T_s$.
In a recent work on detwinned FeSe \cite{TChen2019}, which has no static AF
order below $T_s$ \cite{bohmer}, we again find that the superconductivity-induced resonance only appears
at ${\bf Q}_1$. This further supports the notion that orbital order and the nematic phase below $T_s$
induce the energy splitting of the $d_{xz}$ and $d_{yz}$ bands in electron pockets, which in turn modifies the Fermi surface nesting condition and associated spin fluctuations
along the ${\bf Q}_1$ and ${\bf Q}_2$ directions.

In conclusion, our inelastic neutron scattering experiments on
 mechanically detwinned \BFCA\  with $x=0.048$  and 0.054, which has coexisting
AF order and superconductivity, reveal highly anisotropic spin fluctuations with large magnetic
scattering intensity at the AF ordering wave vector ${\bf Q}_1$ and weak scattering at ${\bf Q}_2$
at temperatures below the
tetragonal-to-orthorhombic structural transition $T_s$.
On cooling to a temperature above $T_c$ but below $T_N$,
a large spin gap appears at the ${\bf Q}_2$ point and spin fluctuations are
mostly centered at the ${\bf Q}_1$ point.
Upon entering the superconducting state, a neutron spin resonance appears
at the ${\bf Q}_1$ point with no magnetic scattering at the ${\bf Q}_2=(0,1)$ point.
By comparing these results with those from ARPES experiments, we conclude that
the anisotropic shift of the $d_{yz}$ and $d_{xz}$ electron-like bands in detwinned \BFCA\ below $T_s$ is associated with the spin excitation anisotropy, and the
superconductivity-induced resonance arises from itinerant electrons with
the $d_{yz}$ orbital characters.  Therefore, low-energy spin fluctuations in underdoped \BFCA\ are highly
orbital selective below $T_s$, suggesting that the orbital order and the nematic phase are correlated with spin fluctuations and superconductivity in underdoped iron pnictide superconductors.

\section{Acknowledgments}
We are grateful to Ming Yi for helpful discussions and detailed explanation of the
ARPES results on {\BFCA}. We thank Yu Song for a critical reading of the paper and
Shiliang Li for allowing us to his laboratory's equipment where
some of the single crystals were grown.
The work at Beijing Normal University is supported by the Fundamental Research Funds for the Central Universities (Grant No. 310432101 and 2014JJCB27) and the National Natural Science Foundation of China (Grant No. 11734002).
The neutron scattering work at Rice is supported by the U.S. NSF-DMR-1700081 (P.D.).

\end{document}